# Generation and removal of apparent singularities in linear ordinary differential equations with polynomial coefficients

S.Yu. Slavyanov[1], D.A. Satco[1], A.M. Ishkhanyan[2,3] and T.A. Rotinyan[1]

[1]Saint Petersburg State University, Saint Petersburg, 198504 Russia
[2]Institute for Physical Research, NAS of Armenia, Ashtarak, 0203 Armenia
[3]Institute of Physics and Technology, National Research Tomsk Polytechnic University, Tomsk, 634050 Russia

**Abstract:** We discuss several examples of generating apparent singular points as a result of differentiating particular homogeneous linear ordinary differential equations with polynomial coefficients and formulate two general conjectures on the generation and removal of apparent singularities in arbitrary Fuchsian differential equations with polynomial coefficients. We consider a model problem in polymer physics.

## 1. Introduction

During recent years, several publications have been devoted to investigating specific properties of ordinary differential equations with apparent singular points (see, e.g., [1-5]). In particular, the important role of these singularities in generating the six Painlevé equations from the Heun equations was revealed in [1]. Equations with apparent singularities were used to expand solutions of the Heun equations in terms of the ordinary and generalized hypergeometric functions in [4,5].

Among the linear ordinary differential equations, those with polynomial coefficients, i.e., equations for which the coefficients of the derivatives of the dependent variable are polynomials in the (complex) independent variable, are rather conspicuous. The zeros of the polynomial at the highest-order derivative of the dependent variable and, in general, the point at infinity are singular points (singularities) of the equation. To examine infinity as a possible singular point, a Möbius transformation can be applied to move the point $z = \infty$ into the finite part of the complex $z$-plane.

If the solutions of a linear differential equation in the vicinity of a finite singular point are functions of temperate growth, i.e., they evolve more slowly than some power function of the distance from the singularity, then the singularity is of the Fuchsian type. The equations with only Fuchsian singularities (including the point at infinity) are called Fuchsian equations. We note that there are Fuchsian equations with nonpolynomial coefficients but which are reducible to such by gauge transformations. On the other hand, gauge transformations are inapplicable to Fuchsian equations with polynomial coefficients because they change the polynomial form of the coefficients.

Here, we show that apparent singularities in linear ordinary differential equations with polynomial coefficients can be generated by differentiating equations that do not involve such singular points and that, inversely, integrating equations containing apparent singularities can result in removing these singularities.

We start with ordinary differential equations with polynomial coefficients for which the top-degree polynomial is at the top-order derivative of the dependent variable. In addition, we suppose that the degree of this polynomial is greater than the order of the differential equation. Hence, we consider the equation

$$\sum_{k=0}^{n} P_k(z) \frac{d^{n-k} w(z)}{d z^{n-k}} = 0, \qquad (1)$$

where $P_0(z)$ is the top-degree polynomial among all $P_k(z)$ and the degree $m$ of this polynomial is greater than the order $n$ of the differential equation: $m > n$. Finally, we suppose that the zeros of polynomial $P_0(z)$ and the point $z = \infty$ (because of some additional restrictions that we do not discuss here) are regular singularities and Eq. (1) is hence Fuchsian.

Discussing several particular examples of such Fuchsian equations, we show that if the polynomial $P_n(z)$ in the last term of the sum in Eq. (1), i.e., the coefficient of the term with w(z), vanishes at a finite point q of the complex plane, then the equation satisfied by the derivative of a solution of Eq. (1) in general involves an additional apparent singularity at this point. Based on this observation, we formulate two general conjectures about generating and removing apparent singular points in arbitrary Fuchsian differential equations with polynomial coefficients.

Further, we discuss confluent reductions of Eq. (1) that contain irregular singularities. We note that irregular singularities are generated by merging the regular singularities via a coalescence procedure. In this case, the coefficients of the resulting equation remain polynomials, but the alternation of the degrees of the polynomials involved in (1) becomes more irregular. We show that generating and removing apparent singularities by respectively differentiating and integrating also work in the case of confluent equations. As an example, we consider a particular model in polymer physics.

## 2. Examples

**Example 2.1.** We consider the general Heun equation which has three Fuchsian singularities at finite points $z_j$, $j = 1,2,3$, and another regular singularity at infinity (without loss of generality one may put $z_1 = 0$, $z_2 = 1$, $z_3 = t$):

$$P_0(z)w''(z) + P_1(z)w'(z) + P_2(z)w(z) = 0, \tag{2}$$

$$P_0(z) = \prod_{j=1}^{3}(z - z_j), \tag{3}$$

$$P_1(z) = \sum_{k=1}^{3}(1 - \theta_k)P_0(z)/(z - z_k), \tag{4}$$

$$P_2(z) = \alpha\theta_\infty(z - q), \tag{5}$$

and the relevant parameters satisfying the Fuchsian identity, which is written as

$$\sum_{k=1}^{3}\theta_k + \theta_\infty + \alpha = 2. \tag{6}$$

The generalized Riemann symbol (see [6]) for this equation is

$$\begin{pmatrix} z_1 & z_2 & z_3 & \infty & z \\ 0 & 0 & 0 & \alpha & q \\ \theta_1 & \theta_2 & \theta_3 & \theta_\infty & \end{pmatrix} \tag{7}$$

We differentiate Eq. (2) with respect to $z$, set $u(z) = w'(z)$, and eliminate $w(z)$ from the resultant equation. We then obtain the equation

$$P_0(z)u''(z) + \left(P_1(z) + P_0'(z) - \frac{P_0(z)}{z-q}\right)u'(z) + \left(P_2(z) + P_1'(z) - \frac{P_1(z)}{z-q}\right)u(z) = 0. \tag{8}$$

This equation, which we call the deformed Heun equation, in addition to the singularities present in the original Heun equation (2), in general has an additional apparent singular point $z = q$. The additional singularity appears if $q$, which is the root of $P_2(z)$, is not a root of $P_0(z)$ [4]. We note that the degree of $P_2(z)$ is equal to unity here, which is the key condition for this result (the case of higher-degree polynomials is considered below). The zeroth-degree case, which is encountered for the hypergeometric equations, does not lead to the appearance of an apparent singularity. Nevertheless, equations in the hypergeometric class with an added apparent singular point are also considered in the literature [3]. Such a situation is achieved by applying additional transformations to the starting equation.

Multiplying now Eq. (8) by $z - q$, we obtain a Fuchsian equation with polynomial coefficients but already with one more finite Fuchsian singular point, and the new polynomial at $u(z)$ is of the second degree. We note that $z = q$ is not a root of this polynomial in the general case. Differentiating Eq. (8) once more, we obtain an equation, generally speaking, with two apparent singular points. Inversely, integrating the equation, we obtain an equation without an apparent singularity.

Concluding this example, we note that instead of the linear Fuchsian second-order differential equations, 2×2 Fuchsian systems of linear first-order equations are often considered; they are reducible to equivalent second-order equations, and the resulting equations contain apparent singular points [7,8].

**Example 2.2**. We again consider a second-order equation of type (2) but with $m$ finite Fuchsian singularities:

$$P_0(z) = \prod_{j=1}^{m}(z - z_j), \tag{9}$$

$$P_1(z) = \sum_{k=1}^{m}(1 - \theta_k) P_0(z)/(z - z_k), \tag{10}$$

$$P_2(z) = \alpha\, \theta_\infty \prod_{j=1}^{m-2}(z - q_j). \tag{11}$$

In this case, the polynomial $P_2(z)$ is of degree $m - 2$, and the Fuchsian identity becomes

$$\sum_{k=1}^{m}\theta_k + \theta_\infty + \alpha = m - 1. \tag{12}$$

We differentiate Eq. (2) with respect to $z$. For the derivative $u(z) = w'(z)$, we then obtain

$$P_0(z)u''(z) + \left(P_1(z) + P_0'(z) - P_0(z)\left(\sum_{j=1}^{m-2}\frac{1}{z - q_j}\right)\right)u'(z) + \left(P_2(z) + P_1'(z) - P_1(z)\left(\sum_{j=1}^{m-2}\frac{1}{z - q_j}\right)\right)u(z) = 0. \tag{13}$$

In the general case where the parameters $q_j$ are different (i.e., $P_2(z)$ does not have multiple roots) and are not the roots of $P_0(z)$, this equation has $m - 2$ additional apparent singular points (the differences between the characteristic exponents is two for each of these points) compared with the original equation.

**Example 2.3.** Let $m = 4$ and $q_1 = q_2 = q$ in the previous example, i.e., $P_2(z) = \alpha\, \theta_\infty (z - q)^2$. Then the point $z = q$ is a double root. By performing essentially the same operations as above, we again obtain an apparent singularity but with different exponents compared with the preceding case. Indeed, in this case,

$$\frac{P_2'(z)}{P_2(z)} = \frac{2}{z-q}, \tag{14}$$

and for the derivative $u(z) = w'(z)$, we hence obtain the deformed equation

$$P_0(z)u''(z) + \left[P_1(z) + P_0'(z) - \frac{2}{z-q}P_0(z)\right]u'(z) + \left[P_2(z) + P_1'(z) - \frac{2}{z-q}P_1(z)\right]u(z) = 0. \tag{15}$$

Accordingly, the difference between the characteristic exponents for $z = q$ is not two, as in the preceding cases, but three. This means that one linearly independent solution of this equation has a zero third derivative at $z = q$, while the power series expansion of the other solution starts from the third degree.

In exactly the same way, we can obtain apparent singular points with the differences between the characteristic exponents equal to four, five, etc.

**Example 2.4.** We consider a particular third-order Fuchsian equation with finite regular singularities located at the points $z_1 = 0$, $z_2 = 1$, $z_3 = t$:

$$z^2(z-1)(z-t)w'''(z) + \left[(3-\alpha-\beta)z(z-1)(z-t) - \theta_2 z^2(z-t) - \theta_3 z^2(z-1)\right]w''(z) + \\ (\alpha-1)(\beta-1)(z-1)(z-t)w'(z) + \kappa(z-q)w(z) = 0. \tag{16}$$

The generalized Riemann symbol representing the characteristic exponents of the singularities of this equation is

$$\begin{pmatrix} z_1 & z_2 & z_3 & \infty & z \\ 0 & 0 & 0 & a & q \\ \alpha & 1 & 1 & b & \\ \beta & 2+\theta_2 & 2+\theta_3 & c & \end{pmatrix} \tag{17}$$

The characteristic exponents in this symbol indicate that the singular point $z = 0$ has one holomorphic solution and two solutions that are generally not holomorphic. At the same time, in the vicinity of $z = 1$ and $z = t$, there are two linearly independent holomorphic solutions and a solution that is not holomorphic. We can indicate some relations between the characteristic exponents at the singular point $z = \infty$ and the parameters $\alpha, \beta, \kappa, \theta_2, \theta_3$. These relations are

$$a + b + c = -\alpha - \beta - \theta_2 - \theta_3, \tag{18}$$

$$ab+bc+ac = \alpha\beta+\theta_2+\theta_3, \tag{19}$$

$$abc = \kappa. \tag{20}$$

Differentiating Eq. (16), we obtain a third-order equation for $u(z) = w'(z)$ with an apparent singularity located at the point $z = q$:

$$u''' + \Delta_1(z)u'' + \Delta_2(z)u' + \Delta_3(z)u = 0, \tag{21}$$

where we introduce the notations

$$\Delta_1(z) = \frac{5-\alpha-\beta}{z} - \frac{\theta_2-1}{z-1} - \frac{\theta_3-1}{z-t} - \frac{1}{z-q}, \tag{22}$$

$$\Delta_2(z) = \frac{(2-\alpha)(2-\beta)}{z^2} + \frac{3-\alpha-\beta}{z}\Sigma(z) - \frac{2\theta_2}{z(z-1)} - \frac{2\theta_3}{z(z-t)} - \frac{\theta_2+\theta_3}{(z-1)(z-t)} + \frac{\theta_2}{(z-q)(z-1)} + \frac{\theta_3}{(z-q)(z-t)}, \tag{23}$$

$$\Delta_3(z) = \frac{(1-\alpha)(1-\beta)}{z^2}\Sigma(z) + \frac{\kappa(z-q)}{z^2(z-1)(z-t)} \tag{24}$$

with

$$\Sigma(z) = \frac{1}{z-1} + \frac{1}{z-t} - \frac{1}{z-q}. \tag{25}$$

This equation can be viewed as a deformed version of Eq. (16).

## 3. Conjectures

The considered examples allow formulating the following conjectures.

**Conjecture 1.** *Any Fuchsian equation with polynomial coefficients having apparent singularities can be obtained from the corresponding derivative of a Fuchsian equation with polynomial coefficients without apparent singularities.*

We note that specifying some parameters is allowed in the course of the calculations. If we use the inverse differentiation, then we obtain the opposite conjecture.

**Conjecture 2.** *Any Fuchsian equation with polynomial coefficients having apparent singularities can be converted into a Fuchsian equation with polynomial coefficients without apparent singularities using the inverse differentiation (integration).*

The proof of these conjectures is an open problem. The main problem is to construct a general concrete representation of a Fuchsian equation with polynomial coefficients.

## 4. Confluent equations

We also apply the differentiation mechanism for generating apparent Fuchsian singularities to confluent equations with polynomial coefficients. But an essentially larger diversity of cases to consider arises here. The problem is that even a formal form of the solution in the vicinity of an irregular singular point is unknown to us, as is also unknown the alternation of the polynomial degrees. We therefore restrict the discussion here to only the simplest equations (but important for applications), namely, the equations of the Heun class.

The confluent Heun equations are second-order linear differential equations that can be represented in the same form as the general Heun equation:

$$P_0(z)w''(z) + P_1(z)w'(z) + P_2(z)w(z) = 0, \qquad (26)$$

but with rules determining the degrees of the involved polynomials that differ from those for Eq. (2). We restrict ourselves to only non-reduced equations, i.e., those for which the power series in the vicinity of the singular points involve only integer powers. In this case, the polynomial $P_0(z)$ is of the second or lower degree, $P_1(z)$ is a second-degree polynomial, and, which is cardinally important, $P_2(z)$ is of the first degree and hence has a zero at one point, for example, at $z = q$. All the involved polynomials depend on the parameter $t$, which passes from one equation to another at the confluence process.

We introduce the derivative $u(z) = w'(z)$. Differentiating Eq. (26) leads to an equation for $u(z)$ that formally coincides with Eq. (8):

$$P_0(z)u''(z) + \left(P_1(z) + P_0'(z) - P_0(z)\frac{P_2'(z)}{P_2(z)}\right)u'(z) + \left(P_2(z) + P_1'(z)) - P_1(z)\frac{P_2'(z)}{P_2(z)}\right)u(z) = 0. \quad (27)$$

Because $P_2(z) = \alpha(z-q)$ for all four confluent Heun equations, in addition to the singularities of the original equations, this equation has an apparent singularity at $z = q$. Naturally, this result holds under the condition that the point $z = q$ does not coincide with any of the already existing singularities, i.e., the apparent singularity appears if $q$ is not a root of $P_0(z)$ [9].

## 6. A physical model

We consider a model in polymer physics. The relaxation time of a polymeric molecule located in a planar hydrodynamic stretching stream can be estimated by solving the spectral problem [10]

$$z(z-1)w''(z)+\left(-\kappa z(z-1)+\frac{3}{2}(z-1)+(b+1)z\right)w'(z)+((\nu-\kappa)(z-1)-2b\kappa z)w(z)=0, \quad (28)$$

where
$$w(0)=0, \quad w(1)=0. \quad (29)$$

The parameter $b$ is called flexibility. It takes large values ($b \sim 10^2 - 10^4$) and determines the size of the region in the vicinity of the Fuchsian singularity $z=1$ where the function $w(z)$ is negligible. The parameter $k$ is expressed as $k=bW$, where $W$ is called the Weissenberg number. This number measures the level of stretching. According to physical representations, the critical value $W=1/2$ marks the coil-stretch transition in elongational flows. For $W<1/2$ polymers are in the coiled state; for $W>1/2$, i.e., if $W$ is greater than the critical value, polymers are fully extended. Finally, the quantity $\nu$ can be regarded as an eigenvalue, and the first eigenvalue $\nu_1$ gives the estimate of the relaxation time

$$T_{rel}=\frac{b\tau}{\nu_1}, \quad (30)$$

where $\tau$ is the time of relaxation of the polymer to equilibrium with the surrounding environment. This problem has been studied numerically many times. In connection with the subject of this paper, we present the equation for the derivative $u(z)=w'(z)$. We introduce the position of the apparent singularity

$$z=q=\frac{\nu-\kappa}{\nu-\kappa-2b\kappa}. \quad (31)$$

Then

$$z(z-1)u''(z)+\left(-\kappa z(z-1)+3(z-1)/2+(b+1)z-\frac{z(z-1)}{z-q}\right)u'(z)+$$
$$\left((\nu-\kappa)(z-1)-2b\kappa z-\kappa(2z-1)+\left(\frac{5}{2}+b\right)-\frac{-\kappa z(z-1)+3(z-1)/2+(b+1)z}{z-q}\right)u(z)=0. \quad (32)$$

This equation can be further used to derive a dynamical nonlinear equation according to the recipe considered in [1].

## 6. Discussion

A specific property of the apparent singularity that makes it distinguishable is as follows. At any ordinary point of a differential equation the Cauchy conditions can be applied; and then all other derivatives can be calculated using the differential equation. In the case of an apparent singularity the Cauchy conditions are still applied, however, the second- or a higher-order derivative can be arbitrary.

A specific property of the apparent singularity distinguishing it is as follows. The Cauchy conditions can be applied at any regular point of a differential equation, and all other derivatives can then be calculated using the differential equation. In the case of an apparent singularity, the Cauchy conditions can still be applied, but the second- or a higher-order derivative can be arbitrary.

The behavior of solutions of linear ordinary differential equations in the vicinity of a Fuchsian singular point is characterized by Frobenius exponents at that point. For the equations with polynomial coefficients, one of these exponents at any finite point is equal to zero. The solution with this exponent is holomorphic in the vicinity of that singularity. A question arises if there exist other holomorphic solutions. The singularities for which such solutions exist are called apparent singularities. The necessary condition for a singularity to be apparent is that the difference between two characteristic exponents is an integer greater than unity.

Here, we have investigated the relations between the equations with and without apparent singular points using several representative examples. We have demonstrated that apparent singularities can be generated in differential equations with polynomial coefficients, both Fuchsian and confluent, as a result of differentiating equations without such singularities. Inversely, apparent singularities can be removed by integration. We have conjectured that these are general properties for Fuchsian equations with polynomial coefficients.

We conclude by noting that using apparent singularities can stimulate several new useful developments. In particular, it allows viewing the Riemann–Hilbert problem in a new way.

**Acknowledgments**

The research of A.M. Ishkhanyan was supported by the Armenian State Committee of Science (SCS Grant No. 15T-1C323) and the project "Leading Russian Research Universities" (Grant No. FTI_24_2016 of the Tomsk Polytechnic University).

**References**

1. S.Yu. Slavyanov and O.L. Stesik, "Antiquantization of deformed Heun-class equations", Theor. Math. Phys. **186**, 118-125 (2016).
2. A.V. Shanin and R.V. Craster, "Removing false singular points as a method of solving ordinary differential equations", Eur. J. Appl. Math, **13**, 617-639 (2002).
3. A.Ya. Kazakov, "Monodromy of Heun equations with apparent singularities", Int. J.

   Theor. Math. Phys. **3**, 190-196 (2013).
4. A. Ishkhanyan and K.A. Suominen, "New solutions of Heun's general equation", J. Phys. A **36**, L81-L85 (2003).
5. C. Leroy and A.M. Ishkhanyan, "Expansions of the solutions of the confluent Heun equation in terms of the incomplete Beta and the Appell generalized hypergeometric functions", Integral Transforms and Special Functions **26**, 451-459 (2015).
6. S.Yu. Slavyanov and W. Lay, *Special functions* (Oxford University Press, Oxford, 2000).
7. A.Ya. Kazakov and S.Yu. Slavyanov, "Euler integral symmetries for a deformed Heun equation and symmetries of the Painlevé PVI equation", Theor. Math. Phys. **155**, 722-733 (2008).
8. S.Yu. Slavyanov, "Polynomial degree reduction of a Fuchsian 2×2 system", Theor. Math. Phys. **182**, 182-188 (2015).
9. V.A. Shahnazaryan, T.A. Ishkhanyan, T.A. Shahverdyan, and A.M. Ishkhanyan, "New relations for the derivative of the confluent Heun function", Armenian J. Phys. **5**, 146 (2012).
10. D. Vincenzi and E. Bodenschatz, "Single polymer dynamics in elongational flow and the confluent Heun equation", J. Phys. A **39**, 10691-10701 (2006).